\documentclass[a4paper]{jpconf}
\usepackage{graphicx}
\usepackage{float}
\usepackage{footmisc}
\usepackage[
  locale=US,
  separate-uncertainty=true,
  per-mode=symbol-or-fraction,
  binary-units,
]{siunitx}

\usepackage{ifthen} 
\newboolean{uprightparticles}
\setboolean{uprightparticles}{false} 
\usepackage{xspace} 
\usepackage{upgreek}

\def\lhcb   {\mbox{LHCb}\xspace}

\def\MagUp {\mbox{\em Mag\kern -0.05em Up}\xspace}

\ifthenelse{\boolean{uprightparticles}}%
{

 \def\PDelta      {\ensuremath{\Delta}\xspace}                 
 \def\PXi         {\ensuremath{\Xi}\xspace}                 
 \def\PLambda     {\ensuremath{\Lambda}\xspace}                 
 \def\PSigma      {\ensuremath{\Sigma}\xspace}                 
 \def\POmega      {\ensuremath{\Omega}\xspace}                 
 \def\PUpsilon    {\ensuremath{\Upsilon}\xspace}

 \def\PB      {\ensuremath{\mathrm{B}}\xspace}                 
                  
 \def\PD      {\ensuremath{\mathrm{D}}\xspace}

 \def\PK      {\ensuremath{\mathrm{K}}\xspace}

 \def\Pi      {\ensuremath{\mathrm{i}}\xspace}

 \def\Ps      {\ensuremath{\mathrm{s}}\xspace}

 \def\thebaroffset{0.0em}
}
{

 \mathchardef\PDelta="7101
 \mathchardef\PXi="7104
 \mathchardef\PLambda="7103
 \mathchardef\PSigma="7106
 \mathchardef\POmega="710A
 \mathchardef\PUpsilon="7107
                  
 \def\PB      {\ensuremath{B}\xspace}                 
                  
 \def\PD      {\ensuremath{D}\xspace}

 \def\PK      {\ensuremath{K}\xspace}

 \def\Pi      {\ensuremath{i}\xspace}

 \def\Ps      {\ensuremath{s}\xspace}

 \def\thebaroffset{0.18em}
}
\newcommand{\offsetoverline}[2][\thebaroffset]{\kern #1\overline{\kern -#1 #2}}%

\makeatletter
\ifcase \@ptsize \relax
  \newcommand{\miniscule}{\@setfontsize\miniscule{4}{5}}
\or
  \newcommand{\miniscule}{\@setfontsize\miniscule{5}{6}}
\or
  \newcommand{\miniscule}{\@setfontsize\miniscule{5}{6}}
\fi
\makeatother

\DeclareRobustCommand{\optbar}[1]{\shortstack{{\miniscule (\rule[.5ex]{1.25em}{.18mm})}
  \\ [-.7ex] $#1$}}

\def\squark    {{\ensuremath{\Ps}}\xspace}

\def\KorKbar {\kern \thebaroffset\optbar{\kern -\thebaroffset \PK}{}\xspace}

\def\D       {{\ensuremath{\PD}}\xspace}

\def\DorDbar {\kern \thebaroffset\optbar{\kern -\thebaroffset \PD}\xspace}

\def\Dp      {{\ensuremath{\D^+}}\xspace}
\def\Dm      {{\ensuremath{\D^-}}\xspace}

\def\DpDm    {\ensuremath{\Dp {\kern -0.16em \Dm}}\xspace}

\def\B       {{\ensuremath{\PB}}\xspace}

\def\BorBbar {\kern \thebaroffset\optbar{\kern -\thebaroffset \PB}\xspace}

\def\Bd      {{\ensuremath{\B^0}}\xspace}

\def\BdorBdbar {\kern \thebaroffset\optbar{\kern -\thebaroffset \Bd}\xspace}

\def\Bs      {{\ensuremath{\B^0_\squark}}\xspace}

\def\BsorBsbar {\kern \thebaroffset\optbar{\kern -\thebaroffset \Bs}\xspace}

\def\Y#1S{\ensuremath{\PUpsilon{(#1S)}}\xspace}

\def\LorLbar     {\kern \thebaroffset\optbar{\kern -\thebaroffset \PLambda}\xspace}

\def\to                 {\ensuremath{\rightarrow}\xspace}

\def\AT#1     {\ensuremath{A_{\mathrm{T}}^{#1}}\xspace}           

\def\C#1      {\ensuremath{\mathcal{C}_{#1}}\xspace}                       
\def\Cp#1     {\ensuremath{\mathcal{C}_{#1}^{'}}\xspace}                    
\def\Ceff#1   {\ensuremath{\mathcal{C}_{#1}^{\mathrm{(eff)}}}\xspace}        
\def\Cpeff#1  {\ensuremath{\mathcal{C}_{#1}^{'\mathrm{(eff)}}}\xspace}       
\def\Ope#1    {\ensuremath{\mathcal{O}_{#1}}\xspace}                       
\def\Opep#1   {\ensuremath{\mathcal{O}_{#1}^{'}}\xspace}                    

\newcommand{\aunit}[1]{\ensuremath{\text{\,#1}}}       

\newcommand{\tev}{\aunit{Te\kern -0.1em V}\xspace}
\newcommand{\gev}{\aunit{Ge\kern -0.1em V}\xspace}
\newcommand{\mev}{\aunit{Me\kern -0.1em V}\xspace}
\newcommand{\kev}{\aunit{ke\kern -0.1em V}\xspace}
\newcommand{\ev}{\aunit{e\kern -0.1em V}\xspace}
 
\newcommand{\mevc}{\ensuremath{\aunit{Me\kern -0.1em V\!/}c}\xspace}
\newcommand{\gevc}{\ensuremath{\aunit{Ge\kern -0.1em V\!/}c}\xspace}
\newcommand{\mevcc}{\ensuremath{\aunit{Me\kern -0.1em V\!/}c^2}\xspace}
\newcommand{\gevcc}{\ensuremath{\aunit{Ge\kern -0.1em V\!/}c^2}\xspace}

\def\ms   {\ensuremath{\aunit{ms}}\xspace}

\newcommand{\chisq}{\ensuremath{\chi^2}\xspace}

\def\gsim{{~\raise.15em\hbox{$>$}\kern-.85em
          \lower.35em\hbox{$\sim$}~}\xspace}
\def\lsim{{~\raise.15em\hbox{$<$}\kern-.85em
          \lower.35em\hbox{$\sim$}~}\xspace}

\def\tell1  {TELL1\xspace}
\def\ukl1   {UKL1\xspace}

\DeclareSIUnit\clight{\text{\ensuremath{c}}}

\begin{document}
\title{Development of the Topological Trigger for LHCb Run 3}
\author{Nicole Schulte$^1$\footnote{Equal contribution} Blaise Raheem Delaney$^{2,3}$\footnotemark[\value{footnote}], Niklas Nolte$^{2,3}$\footnotemark[\value{footnote}], Gregory Max Ciezarek$^4$, Johannes Albrecht$^1$, Mike Williams$^{2,3}$}

\ead{nicole.schulte@cern.ch}

\address{$^1$ TU Dortmund University, Dortmund, Germany}
\address{$^2$ Massachusetts Institute of Technology, Cambridge MA, USA}
\address{$^3$ NSF AI Institute for Artificial Intelligence and Fundamental Interactions (IAIFI)}
\address{$^4$ CERN, Meyrin, Switzerland}

\begin{abstract}
The data-taking conditions expected in Run 3 of the LHCb experiment at CERN are unprecedented and challenging for the software and computing systems. Despite that, the LHCb collaboration pioneers the use of a software-only trigger system to cope with the increased event rate efficiently. The beauty physics programme of LHCb is heavily reliant on topological triggers. These are devoted to selecting beauty-hadron candidates inclusively, based on the characteristic decay topology and kinematic properties expected from beauty decays. The following proceeding describes the current progress of the Run 3 implementation of the topological triggers using Lipschitz monotonic neural networks. This architecture offers robustness under varying detector conditions and sensitivity to long-lived candidates, improving the possibility of discovering New Physics at LHCb.
\end{abstract}

\section{Introduction}

In Run 3, the \lhcb experiment \cite{LHCb} operates with a software-based trigger system 
\cite{TDR}, making it one of the first experiments 
to process the incoming data generated in proton-proton and heavy-ion 
collisions without prior hardware selection. This necessitates employing intelligent and fast selection 
algorithms that can efficiently process the data while covering a broad physics programme within the allocated computing resources \cite{TDR}. One of the main 
selection algorithms in LHCb is the so-called topological trigger, aiming to select beauty decays inclusively 
based on their distinct topology \cite{BBDT}\cite{2011trigger}. As beauty decays are one of the main interests in LHCb, it is essential 
to determine this selection algorithm carefully since its 
output is used for most analyses in LHCb. In Run 2, the selection algorithm of the topological triggers was based on 
boosted decision trees \cite{run2trigger}, which were providing around \SI{80}{\percent} 
signal efficiency in total when tested on various decays under the given conditions of the experiment. For Run 3, 
the conditions have changed drastically, with LHCb increasing the luminosity, which causes more recorded primary vertices per event. 
To make use of the increase in available data, LHCb has chosen to upgrade its detector for Run 3.
Not only is the detector hardware in LHCb advancing, but the field of machine learning algorithms also has progressed, opening the possibility to improve the algorithms used for physics research as well. This is why
the topological trigger in Run 3 is based on neural networks that provide robustness against
detector effects and sensitivity to outlier particles that are classified as potentially interesting Beyond the 
Standard Model (BSM) candidates that are not represented in the training data.

\section{The Topological Trigger} 
The LHCb detector is one of the four experiments stationed at the Large Hadron Collider at CERN. 
It is a single-arm forward spectrometer that covers a pseudorapidity range of $2 \textless \eta \textless 5$. The main 
interest of LHCb is the study of heavy-flavour decays with a particular focus on beauty hadrons. In Run 3, the detector 
operates under new conditions as most of the detector parts are upgraded or replaced, and the luminosity has increased 
by a factor of $5$ compared to the previous run of data taking \cite{TDR}. 

\begin{figure}[t]
\includegraphics[width=\textwidth]{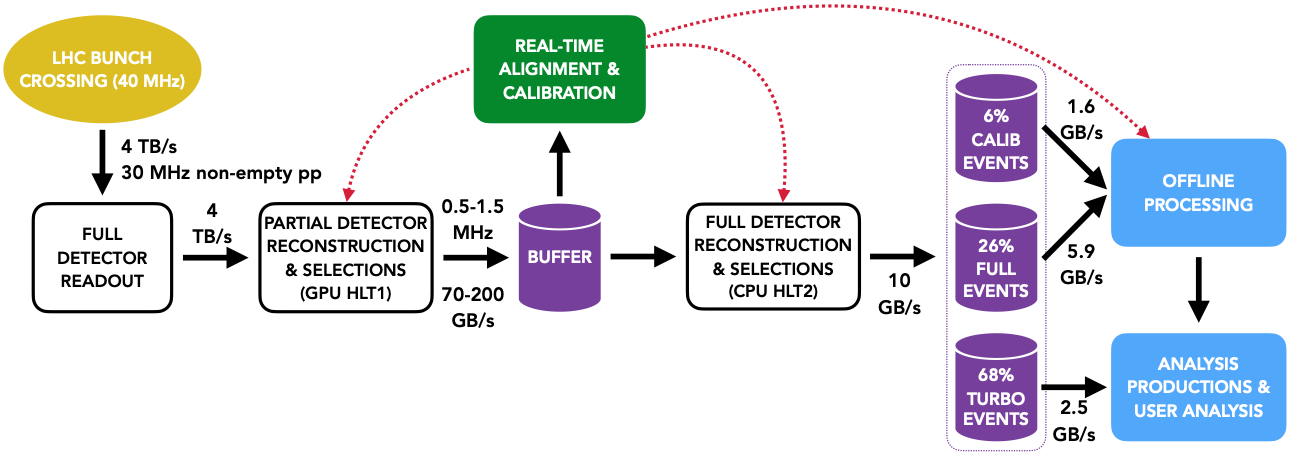}\hspace{2pc}%
\caption{\small{\label{dataflow} Schematic representation of the data flow in LHCb. The incoming rate of \SI{30}{\mega\hertz} of non-empty bunch crossings is processed by the full detector readout and then passed though various stages of the software trigger system before 
ultimately getting stored for analysis production. From \cite{dataflow}.}}
\end{figure}

The change in conditions is challenging for the detector and the trigger system, which operates
fully software-based for the first time ever. The software has to reduce the 
incoming rate of \SI{30}{\mega\hertz} of non-empty proton-proton bunch crossings to a rate of around \SI{100}{\kilo\hertz}, which in turn is passed 
to storage. A schematic overview of the data flow in the LHCb software can be found in Figure \ref{dataflow}. After 
the full detector readout, the incoming data is passed onto the first selection in the trigger system, the 
High Level Trigger 1 (HLT1). Using information from the tracking system and primary vertex information, HLT1 processes 
the data down by a factor of 30. The selected data is then
passed onto a buffer system allowing real-time alignment and calibration to perform the
calculations needed for the High Level Trigger 2 (HLT2) selections.
HLT2 is the second selection step in the trigger system, which has the full detector reconstruction 
available to make more specific selections. Many selection filters are written in this part of the trigger system targeting
a wide range of exclusive and inclusive decays. Inclusiveness in this 
case refers to the selection of an ensemble of decays sharing similar topologies, rather than one 
specific decay process.

The topological trigger produces the largest output bandwidth of any HLT2 selection algorithm. It aims to select beauty decays inclusively,
based on their topology. To this end, the algorithm is trained on various
exclusive decays covering a broad spectrum of the LHCb beauty-physics programme. The resulting model
can thus perform inference to select processes compatible with $b$-hadron decays. Furthermore, each exclusive simulation contributes the same amount of signal candidates to the training. This procedure encourages a non-biased selection of the topological triggers. 
Beauty decays display a distinct signature in LHCb since they are boosted 
in the forward direction of the detector. Due to the relatively high lifetime of around \SI{1.6}{\pico\second} \cite{lifetimes}, a beauty hadron traverses the 
detector up to $\mathcal{O}(\text{cm})$ before decaying. The distance that the particle traverses before decaying is often denoted as the flight distance. The secondary decay vertices can be identified in the 
detector and can be used to distinguish beauty decays from other interactions. Charm decays display a similar topology.
Although charm particles have a lifetime that is four times shorter \cite{lifetimes} than beauty particles, they also traverse $\mathcal{O}(\text{mm})$ before 
decaying. The cross-section of charm decays is $\mathcal{O}(10)$ \cite{lifetimes} times higher than for beauty decays, making charm contributions one of the 
most prominent backgrounds for the topological trigger alongside soft-QCD and 
combinatorial backgrounds.

Two versions of the Run 3 topological trigger are implemented into the \lhcb software stack. These algorithms are trained to target decays with at least 
two or three charged particles, respectively. 
Combined, the topological triggers enable the selection of multi-body beauty decays. 
The topological 
trigger writes to the so-called {\it{Full Stream}}~\cite{TDR},  meaning all the information of the selected events is stored. Therefore, $n$-body $B$ decays, with $n>3$, may be identified as signal in the two- and three-body combinations by the topological triggers.

The topological triggers are run over composite candidates, reconstructed as follows: two reconstructed input 
particles are selected according to minimal criteria on kinematic 
variables like the momentum. Afterwards, a vertex fit is performed to 
infer whether the two particles originate from the same primary vertex. 
The surviving combinations are then considered as a two-body object, which 
is treated as a single particle from this point onward. In the
case of the three-body algorithm, another particle is added to the two-body object forming a three-body candidate, which itself is treated as 
a single particle from this point. These combinations are also filtered 
according to vertex quality and kinematic criteria.

\section{Monotonic Lipschitz Neural Networks} 
Whilst the Run 1 and Run 2 \cite{BBDT}\cite{2011trigger} topological triggers exploited BDTs, the Run 3 implementation makes use of monotonic Lipschitz neural networks \cite{NN}. 
These neural networks are trained to provide robustness against the detector conditions varying throughout data taking by constraining the response of the neural network 
by the Lipschitz constant. 
Furthermore, the Run 3 topological triggers have been developed to increase sensitivity to yet-undiscovered Beyond the Standard Model candidates that are not included in the training data. 

Varying detector conditions can lead to undesired outliers in the neural network response and, consequently, complicate the evaluation of the relevant systematic uncertainties in physics measurements.
Robustness against detector effects is achieved by introducing a constraint on the Lipschitz constant of the neural networks response. To give an example on how the Lipschitz constant 
is used, one can consider two inputs $x$ and $x^\prime$, with each input representing a vector with an entry for each feature used 
for classification together with the classification model $M$. The Lipschitz constant $\lambda$
itself is defined by the distance between the response of the neural network $M(x)$ and 
$M(x^\prime)$ and
the distance between the inputs themselves. Robustness can then be ensured by constraining the Lipschitz constant to an upper value. This effectively ensures that 
detector effects of limited magnitude have a strict upper limit on their effect on the response.

Monotonicity complements the robustness requirement, opening the possibility of selecting
interesting outliers that could potentially be BSM candidates that are not
known prior to the training of the algorithm and are therefore not learned by the classifier.
Consider two data points $x$ and $x\prime$ that differ only in one feature $i$, a model $M$ is  monotonically increasing in feature $i$, if 
$x_i < x{_i}{^\prime}$ implies $M(x) 
< M(x^\prime)$. The architecture allows constraints to be monotonic in either the increasing or decreasing direction for each feature, individually.

\begin{figure}[H]
\includegraphics[width=\textwidth]{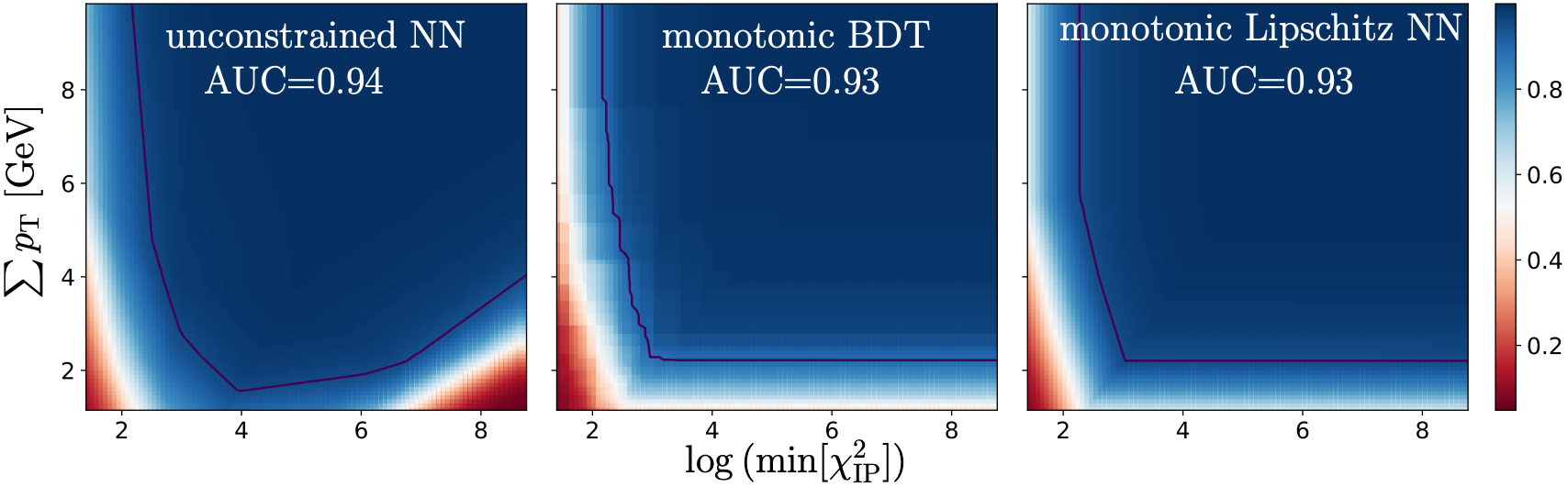}\hspace{2pc}%
\caption{\small{\label{NNarch} Comparison of three classification algorithms on an example classification problem. The 
red area represents events that are classified as background, the blue area represents events that are selected as
signal and the black curve represents the decision boundary. The left panel corresponds 
to the output of an unconstrained NN, the middle panel shows the response of a monotonic 
BDT and the right panel corresponds to the output of a monotonic Lipschitz NN. Taken from \cite{NN}.}}
\end{figure}

Figure \ref{NNarch} displays the advantage of this architecture on an example binary classification task using only two input
features for classification: the sum of the transverse momentum and the 
$\chi^2$ of the impact parameter, which is a measure of displacement of a particle. The presence of background can be seen in the phase space of high displacement and low momentum 
candidates. The conventional neural network is learning to reject this region in the selection and therefore also disfavouring signal candidates with a higher $\chi^2$.
In contrast, the monotonic boosted decision
tree provides more sensitivity towards highly displaced objects, but also an undesired, 
uneven decision boundary, which in turn could cause issues when analysing the 
distribution of the transverse momentum. The monotonic Lipschitz
neural network, on the other hand, is sensitive to the particles in the region of phase space characterised by
higher displacement and low momentum while maintaining a smooth decision boundary throughout.

\section{Application and Performance} 
The topological triggers adopt a feature set comprising various kinematic and spatial features such as transverse 
momentum $p_T$, flight distance $\chi^2$ and $\textrm{IP} \chisq$, and the impact parameter $\chi^2$ with respect to the primary vertex of the multi-body candidates. The feature set has been optimised to capture the topology and momentum budget of beauty decays whilst discriminating against the prompt-charm and combinatorial backgrounds in the event. The background 
for the classifier training is taken from a minimum bias sample. This sample is assembled to represent the average content of a $pp$ collision. After filtering out the beauty events from this 
sample, it is the ideal sample to model the background information considered by the classifier.

The NN response is required to be monotonic with respect to a subset of the features. 
Specifically, the Run 3 topological trigger response must be monotonically increasing with respect to the $p_T$ and $\textrm{IP}\,\chisq$ of multi-body candidates. In this way, an inductive bias is introduced to bolster sensitivity to highly boosted, high-momentum candidates. Such conditions are optimised for the selection of beauty candidates and feebly interacting, long-lived BSM candidates. Additionally, the 
kinematic variables are rescaled to a range of $\mathcal{O}(1)$ to provide the neural network with inputs that are in the same range. To realise this, all kinematic variables are 
scaled in units of $\SI{}{\giga\electronvolt\per\clight}$ and the logarithm of all vertex-fit quality variables is taken. A $5\sigma$ window around the mean of the variable is calculated thereafter, and values that exceed this interval are clipped onto its outer bins.

The Lipschitz constant for the two and three-body algorithms has been optimised independently to achieve a high reconstruction efficiency on signal simulations whilst being compatible with the resolution expected of the \lhcb detector. In loose terms, compatible means that the Lipschitz constants are small enough to disallow significant changes to the classification score when changing inputs within their resolution scale.
A scan of varying values of $\lambda$ has been performed, yielding the values $\lambda=1.75$ and $\lambda =2.0$ for the two- and three-body triggers, respectively.

\begin{figure}[t]
\begin{minipage}{19pc}
\includegraphics[width=19pc]{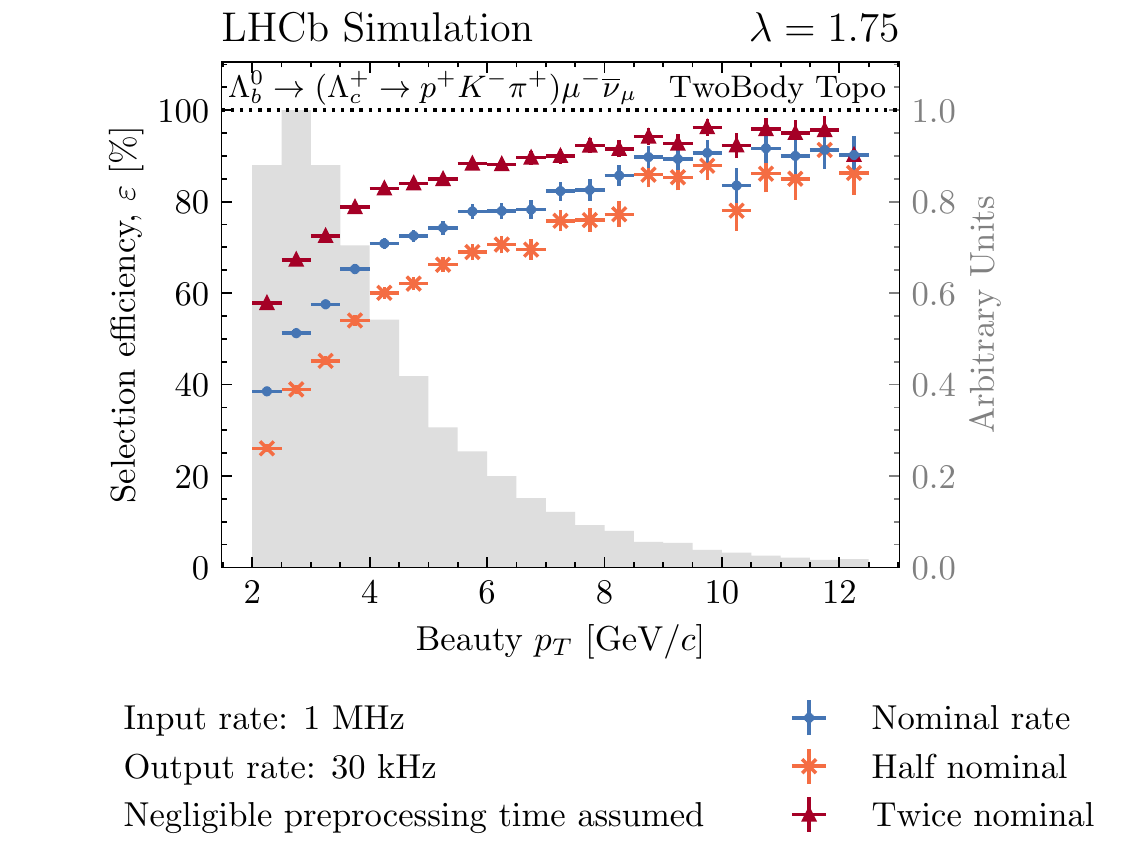}
\end{minipage}%
\begin{minipage}{19pc}
\includegraphics[width=19pc]{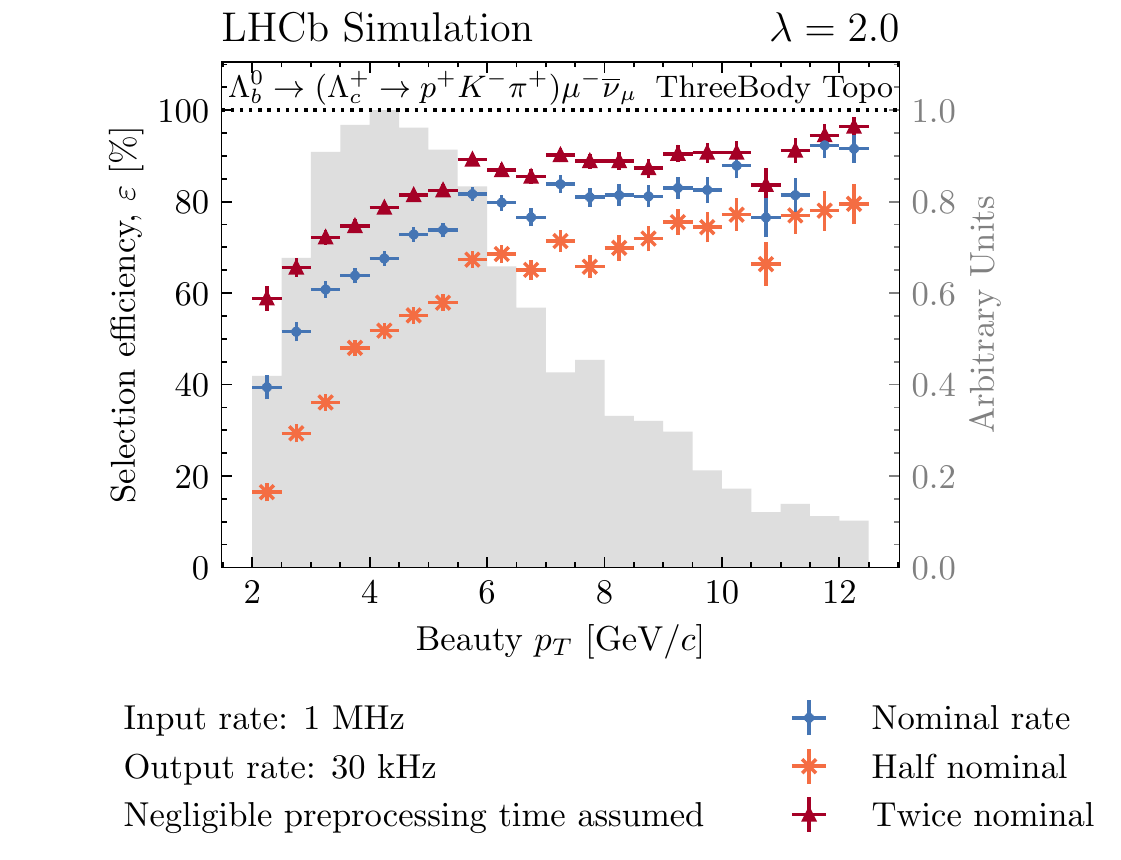}
\end{minipage} 
\caption{\small{\label{performance_LcMuNu} Signal efficiency in units of the transverse momentum for the two-body selection 
(left) and three-body selection (right). The different scenarios are displayed, corresponding to either a 
nominal output rate of the topological trigger of \SI{30}{\kilo\hertz}, twice the nominal output rate or half of it. The 
efficiency is calculated on a Monte Carlo sample for the 
decay $ \Lambda_b^0 \to (\Lambda_c^+ \to p^+ K^- \pi^+) \mu^- \overline{\nu}_\mu $ .}}
\end{figure}

Figure \ref{performance_LcMuNu} shows the signal reconstruction efficiency as a function of beauty transverse momentum, evaluated on simulated 
 $ \Lambda_b^0 \to (\Lambda_c^+ \to p^+ K^- \pi^+) \mu^- \overline{\nu}_\mu $ decays that are excluded from the training set.
As the topological trigger is expected to retain 
most of the HLT2 output bandwidth, it is important to calculate the reconstruction efficiency on signal for different scenarios concerning the amount of output data 
that gets stored per second. This is needed to make an informed decision about the 
computing resources used for these selection algorithms. The output 
bandwidth itself can be calculated by multiplying the output rate by 
the average event size, which is roughly around \SI{100}{\kilo\byte} \cite{TDR}.
Three different bandwidth scenarios for either \SI{1.5}{\giga\byte\per\second}, \SI{3}
{\giga\byte\per\second} and \SI{6}{\giga\byte\per\second} are displayed for each algorithm individually as unofficial working points to 
get a first understanding of the performance of the selection.
It can be seen that even 
with half the nominal rate of \SI{15}{\kilo\hertz}, a signal efficiency of around 
\SI{80}{\percent} for candidates at high transverse momentum can be maintained. 

\begin{figure}[t]
\begin{minipage}{19pc}
\includegraphics[width=19pc]{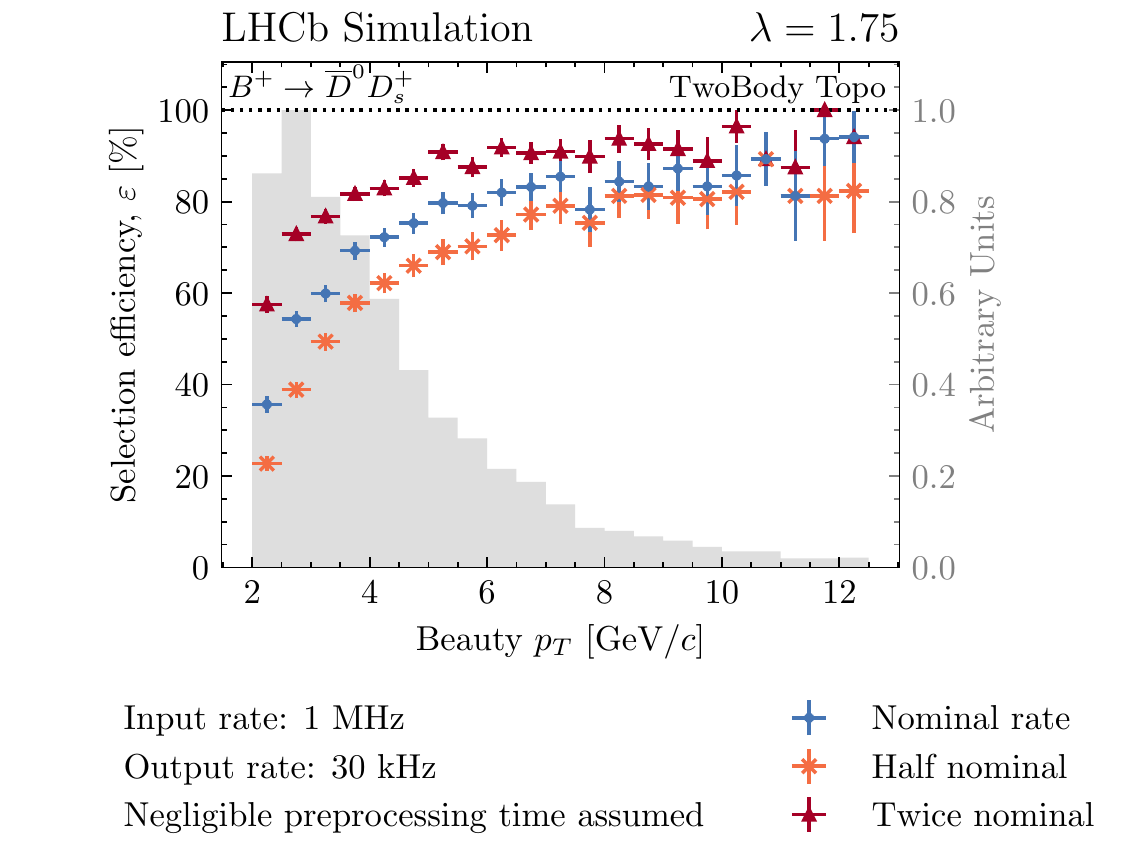}
\end{minipage}%
\begin{minipage}{19pc}
\includegraphics[width=19pc]{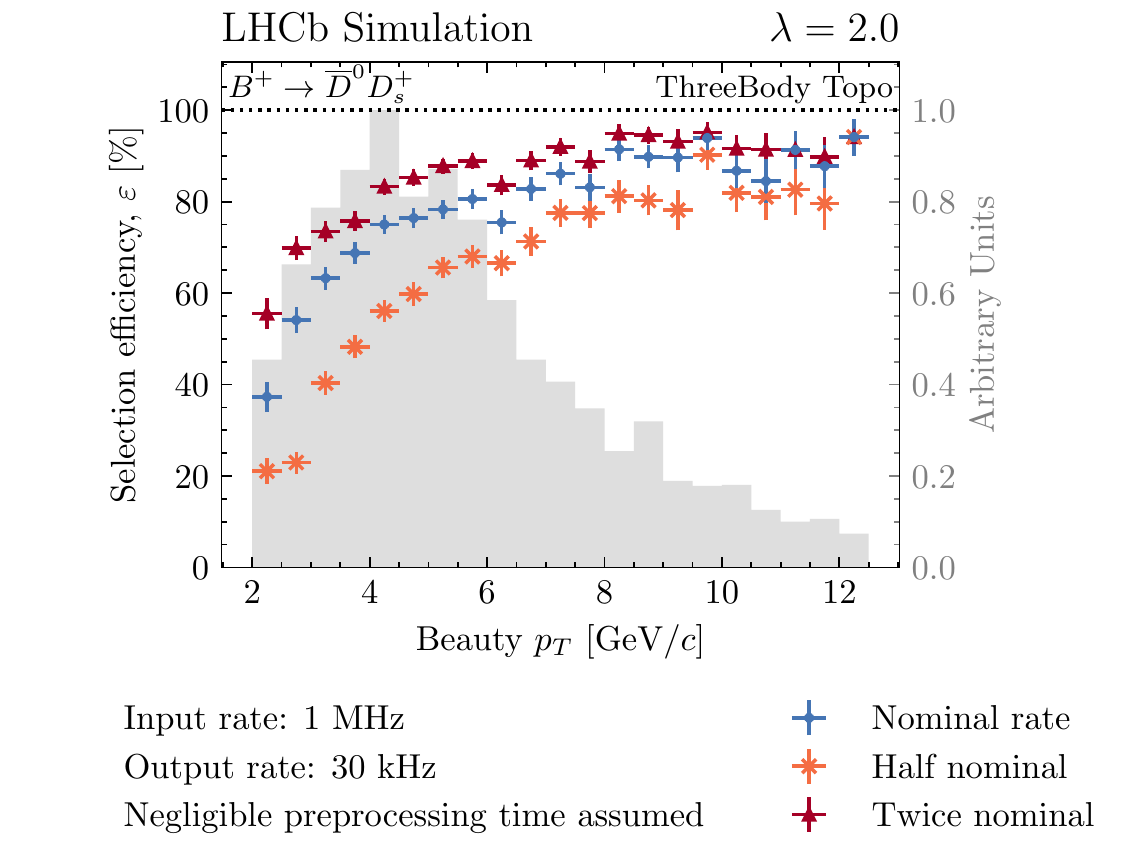}
\end{minipage} 
\caption{\small{\label{performance_D0Ds} Signal efficiency as a function of the transverse momentum for the two-body selection 
(left) and three-body selection (right). The different scenarios are displayed, corresponding to either a 
nominal output rate of the topological trigger of \SI{30}{\kilo\hertz}, twice the nominal output rate or half of it. The 
efficiency is evaluated on a Monte Carlo Sample for the decay $B^+ \to \bar{D}^0 D^+_s$.}}
\end{figure}

Figure \ref{performance_D0Ds} shows the signal efficiency as a function of the transverse momentum for the two-body and 
three-body algorithm on an exclusive Monte Carlo simulation of the decay \mbox{$B^+ \to \bar{D}^0 \left(\to K^+ \pi^- \right) D^+_s \left(\to K^+ K^- \pi^+ \right)$}. Manifestly the three-body trigger delivers a comparatively higher selection efficiency owing to the high multiplicity of the final-state tracks. Broadly, this result demonstrates the capacity to reconstruct beauty decays with high final-state multiplicity.

\section{Conclusion} 
The topological trigger of LHCb is a selection algorithm devoted to selecting  beauty decays inclusively. An inclusive selection is achieved by training the model on a sample amounting to a mixture
of exclusive beauty decays that are equally considered in the training of the classifier. This is done in favour of reducing biases towards any sort of decay. Using a mixture exclusive decays in 
the training stage tasks the classifier to generalise the topology of beauty decays. This 
generalisation enables the selection of decays that are not directly included in the training of the trigger itself.
The implementation of the topological triggers as Lipschitz monotonic neural networks protects the selection against inefficiencies due to detector effects. Furthermore, this class of architectures increases the sensitivity to 
candidates with a higher momentum budget, thereby boosting sensitivity to long-lived beauty and BSM candidates. 
Whilst the optimisation of such triggers in Run 3 is ongoing, a preliminary successful inclusive selection 
of beauty decays has already been demonstrated. 
This contribution demonstrates that the current algorithm is successful 
in the selection of inclusive beauty candidates when evaluated on two simulated probe decay channels that are not included in the training set. For the future, the selection algorithms 
will be refined further and optimised to achieve ideal timing while maximising signal efficiency.

\ack{
The authors would like to thank the LHCb computing and simulation teams for their support and for producing the simulated LHCb samples used in the paper. The authors would also like to thank the LHCb online, DPA and RTA team for providing most of the software that this project was built upon and for all the support and guidance that was given.
NS and JA acknowledge funding from the German Science Foundation DFG, within the Collaborative Research Center SFB1491 "Cosmic Interacting Matters - From Source to Signal." 
BD, NN, and MW were supported by NSF grant PHY-2019786 (The NSF AI Institute for Artificial Intelligence and Fundamental Interactions, http://iaifi.org/).
BD and MW were also supported by U.S.\ NSF grant PHY-2209181. JA acknowledges funding from the German Federal Ministry of Education and Research (BMBF, grant no. 05H21PECL1) within ErUM-FSP T04.}

\section{References}

\end{document}